\documentclass[sigconf]{acmart}

\usepackage{fancyhdr}
\usepackage{subfig}
\usepackage{threeparttable}
\usepackage{multirow}
\usepackage{makecell}
\usepackage{wasysym}
\usepackage[linesnumbered,ruled,vlined]{algorithm2e}
\newcommand{\redsad}{\textcolor{red}{\frownie}}
\newcommand{\grnsml}{\textcolor{teal}{\smiley}}

\AtBeginDocument{%
  }

\copyrightyear{2025}
\acmYear{2025}
\setcopyright{acmlicensed}
\acmConference[ASPDAC '25]{30th Asia and South Pacific Design Automation Conference}{January 20--23, 2025}{Tokyo, Japan}
\acmBooktitle{30th Asia and South Pacific Design Automation Conference (ASPDAC '25), January 20--23, 2025, Tokyo, Japan}
\acmDOI{10.1145/3658617.3697558}
\acmISBN{979-8-4007-0635-6/25/01}


\begin{document}

\title{A Hierarchical Dataflow-Driven Heterogeneous Architecture for Wireless Baseband Processing}

\author{Limin Jiang, Yi Shi, Yintao Liu, Qingyu Deng, Siyi Xu, Yihao Shen, Fangfang Ye, Shan Cao, and Zhiyuan Jiang$^*$}
\affiliation{%
  \country{School of Communication and Information Engineering, Shanghai University, Shanghai, China}
}
\email{jiangzhiyuan@shu.edu.cn}

\renewcommand{\shortauthors}{Jiang et al.}

\begin{abstract}
Wireless baseband processing (WBP) is a key element of wireless communications, with a series of signal processing modules to improve data throughput and counter channel fading. Conventional hardware solutions, such as digital signal processors (DSPs) and more recently, graphic processing units (GPUs), provide various degrees of parallelism, yet they both fail to take into account the cyclical and consecutive character of WBP. Furthermore, the large amount of data in WBPs cannot be processed quickly in symmetric multiprocessors (SMPs) due to the unpredictability of memory latency. To address this issue, we propose a hierarchical dataflow-driven architecture to accelerate WBP. A \textit{pack-and-ship} approach is presented under a non-uniform memory access (NUMA) architecture to allow the subordinate tiles to operate in a bundled access and execute manner. We also propose a multi-level dataflow model and the related scheduling scheme to manage and allocate the heterogeneous hardware resources. Experiment results demonstrate that our prototype achieves $2\times$ and $2.3\times$ speedup in terms of normalized throughput and single-tile clock cycles compared with GPU and DSP counterparts in several critical WBP benchmarks. Additionally, a link-level throughput of $288$ Mbps can be achieved with a $45$-core configuration.
\end{abstract}




\maketitle

\section{Introduction}
As 5G technology continues to expand, network operators are looking for a more energy-efficient and open-source hardware solution to replace the proprietary one offered by device vendors \cite{garcia2021ran}. The hardware implementation of wireless baseband processing (WBP), which is data and computation-intensive, is becoming a major challenge to this end.

A variety of solutions have been proposed to address the need for WBP. Digital signal processors (DSPs) are the first in this regard and are known for their very long instruction word (VLIW) architecture, which allows for instruction-level parallelism and can be used in many digital processing domains. However, the high control overhead associated with VLIW limits its scalability to greater parallelism, which is essential for WBP. More recently, with the rise of general-purpose graphic processing units (GPGPUs), several attempts have been made to take advantage of GPUs' massive computing capability, but their high power consumption has hindered their further use. Similarly, manycore servers have also been proposed to meet the high throughput requirements of WBP, but have failed to overcome the power consumption barrier.

\begin{figure}[!t]
  \centering
  \includegraphics[width=\linewidth]{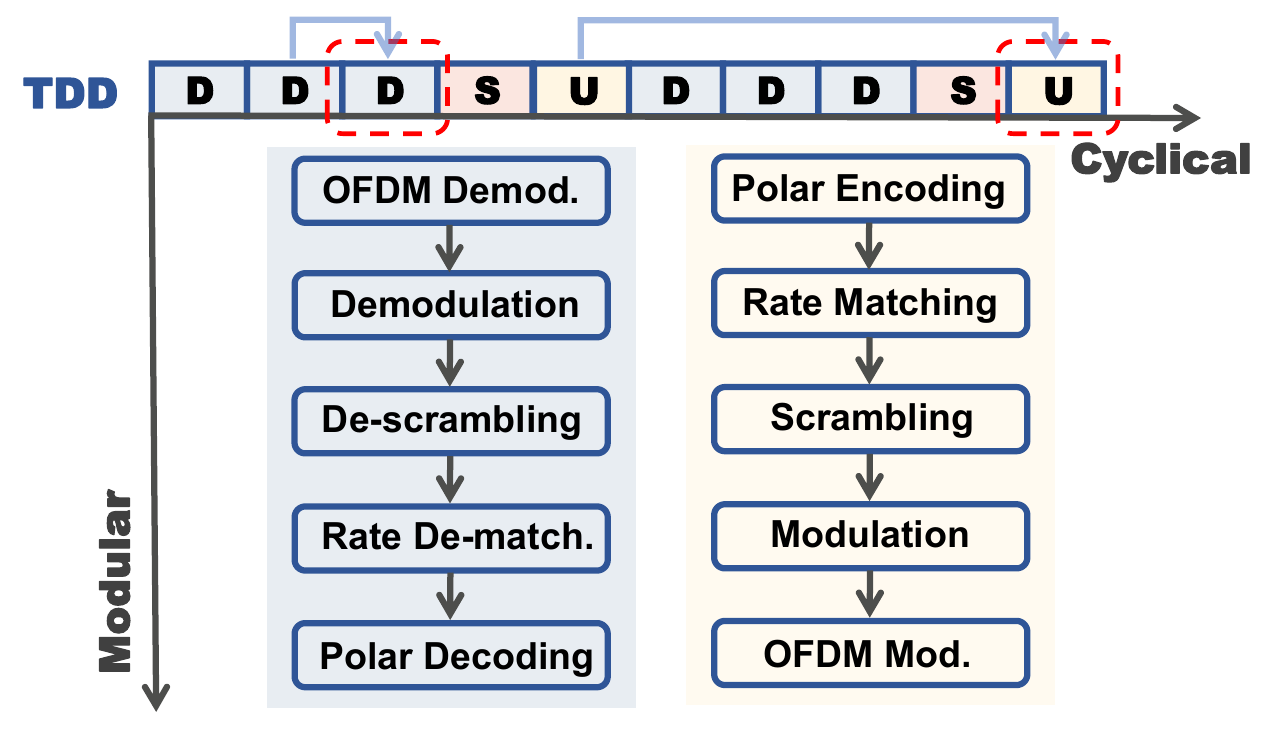}
  \caption{A TDD frame structure in wireless communication (OFDM: Orthogonal frequency division multiplexing).}
  \label{intro}
  \vspace{-0.3cm}
\end{figure}

WBP has two distinct characteristics, cyclical and consecutive (modular). Figure \ref{intro} illustrates a typical time division duplex (TDD) wireless communication frame structure. The uplink and downlink processes can be logically divided into several data-independent and successive modules, for example, scrambling after rate-matching, where data are transferred in a grouped manner and the activation of the following module depends on the completion of the previous output. Additionally, both sides demonstrate cyclical patterns along the temporal axis, decoding or generating signals after a certain period of time regulated by communication protocols.

Taking advantage of the modular and cyclical characteristics of WBP, which is predictable and decoupled in nature, this paper presents a novel hierarchical dataflow-driven architecture. To address the issue of unpredictable memory access and improve data locality, multiple cache-free single instruction multiple data (SIMD) cores are arranged in a non-uniform memory access (NUMA) configuration. Our main contributions are summarized as follows:

\begin{itemize}
    \item A cache-free manycore architecture is proposed to increase energy and area efficiency without compromising performance due to the predictable data processing nature of WBP.
    \item We develop a pack-and-ship data dispatch system to enable the tiles to operate in a bundled access and execution style, which can drastically reduce the cost of data movement.
    \item A hierarchical dataflow task scheduling scheme is designed and two strategies, namely multi-threading and lazy-deletion, are proposed to fully utilize the hardware resources.
\end{itemize}






\section{Related Works}
DSPs remain the most popular choice for WBP due to their quick time-to-market, moderate power, performance and area (PPA) and programming flexibility. The VLIW architecture takes advantage of instruction-level parallelism (ILP) and micro-architectural enhancements are made to address the increasing service requirements. The Hexagon processor \cite{codrescu2014hexagon} introduces dynamic multi-threading, allowing one thread to remain active on the physical core if other threads are idle or stalled, instead of executing in a round-robin fashion. The latest Hexagon V73 processor has extra execution units such as vector and tensor integrated into the pipeline for neural processing \cite{mahurin2023qualocmm}. NXP provides solutions for different base station coverage with StarCore DSPs and IBM Power Architecture cores \cite{QorIQ2013}. Ceva's XC architecture \cite{ceva2020} supports 5G baseband processing with a dynamically configurable multi-thread feature. Scalar processors can be assigned dual or quadruple vector computation units for better hardware utilization. However, proprietary compilers must implement sophisticated optimizations to achieve optimal ILP for DSPs. The emergence of hardware accelerators in system design also suggests the insufficiency of DSPs for future wireless systems.

\begin{table}
  \caption{Classification of related works}
  \label{classfy}
  \begin{threeparttable}[b]
  \begin{tabular}{rccccc}
    \toprule
    Work & Core Het. & Scal. & DLP & TLP & HW/SW \\
    \midrule
    TeraPool \cite{bertuletti2023efficient} & \redsad & \grnsml & \grnsml & -- & \grnsml\\
    DXT501 \cite{chen2022dxt501} & \grnsml & \redsad & \grnsml & \grnsml & \grnsml\\
    MAGALI \cite{jalier2010heterogeneous} & \grnsml & \grnsml & \redsad & -- & \redsad\\
    MACRON \cite{ling2015macron} & \grnsml & \grnsml & \grnsml & -- & \grnsml\\
    Sora \cite{tan2011sora} & \redsad & \redsad \tnote{$\ast$} & \grnsml & \grnsml & \redsad\\
    SPECTRUM \cite{venkataramani2020spectrum} & \redsad & \grnsml & \redsad & \grnsml & \redsad\\
    Ours & \grnsml & \grnsml & \grnsml & \grnsml & \grnsml\\
  \bottomrule
\end{tabular}
\begin{tablenotes}
    \footnotesize
    \item [$\ast$] Runs on a commercial general-purpose processor.
\end{tablenotes}
\end{threeparttable}
\end{table}

Recently, GPUs have become a viable alternative to WBP. Their massive data-level parallelism and self-contained libraries make them well-suited for multi-user multiple-input multiple-output (MU-MIMO) systems. NVIDIA provides a platform that can be used for both 5G and machine learning development \cite{kelkar2021nvidia}. The Aerial software development kit includes cuBB and cuVNF, which help accelerate physical layer and packet processing, respectively. A MIMO testbed is also presented in \cite{kelkar2021nvidia}, which uses a GPU for baseband processing and a data processing unit for fronthaul data transmission. However, the large amount of power dissipated by the numerous compute unified device architecture (CUDA) cores and dedicated rendering-related processing units makes it difficult to deploy on more energy-efficient user equipment.

Various works have been presented in academia seeking a way towards manycore parallel computing for WBP. We evaluate and classify the related works based on the following criteria:
\begin{itemize}
    \item \textbf{Core heterogeneity:} The system as a whole can be made more energy-efficient if the right processor is assigned a specific workload.
    \item \textbf{Scalability:} The hardware design of the system should be able to be scaled to fit the varying throughput needs of wireless communication protocols.
    \item \textbf{Data-level parallelism (DLP):} DLP techniques such as vector processing eliminate redundant instruction fetch and decoding, which is essential in WBP.
    \item \textbf{Thread-level parallelism (TLP):} The overall system should be capable of handling multiple processes (e.g. transmitting and receiving threads) at the same time.
    \item \textbf{Hardware and software (HW/SW) co-design:} Improvements are made in both HW and SW instead of utilizing off-the-shelf processors, compilers and algorithms.
\end{itemize}

Table \ref{classfy} presents some of the representative architectures for WBP. In \cite{ding2020agora}, \cite{tan2011sora} and \cite{yang2013bigstation}, software scheduling algorithms and optimizations are implemented on commercial enterprise-grade general-purpose processors (GPPs). To achieve optimal core utilization, multiple dimensions of parallelism are exploited. However, the versatility of the core is a disadvantage since many ILP techniques such as out-of-order and speculative execution have limited performance gains on domain-specific computing. Although there are a few manycore architectures \cite{bertuletti2023efficient, schoeberl2015t, venkataramani2020spectrum} along with task-mapping methodologies or worst-case execution time analyses to guarantee the time-predictability of the system, they overlook core heterogeneity and, furthermore, the methodologies are not easily transferable to heterogeneous platforms. In contrast, both \cite{jalier2010heterogeneous} and \cite{ling2015macron} map tasks on heterogeneous network-on-chip frameworks, but the lack of DLP and the chip interconnect impede their performance, respectively. An architecture containing application-specific instruction set processors (ASIPs) is proposed in \cite{chen2022dxt501}, but it failed to scale up the ASIP cores due to the large die size.

\begin{figure}[!t]
  \centering
  \includegraphics[width=\linewidth]{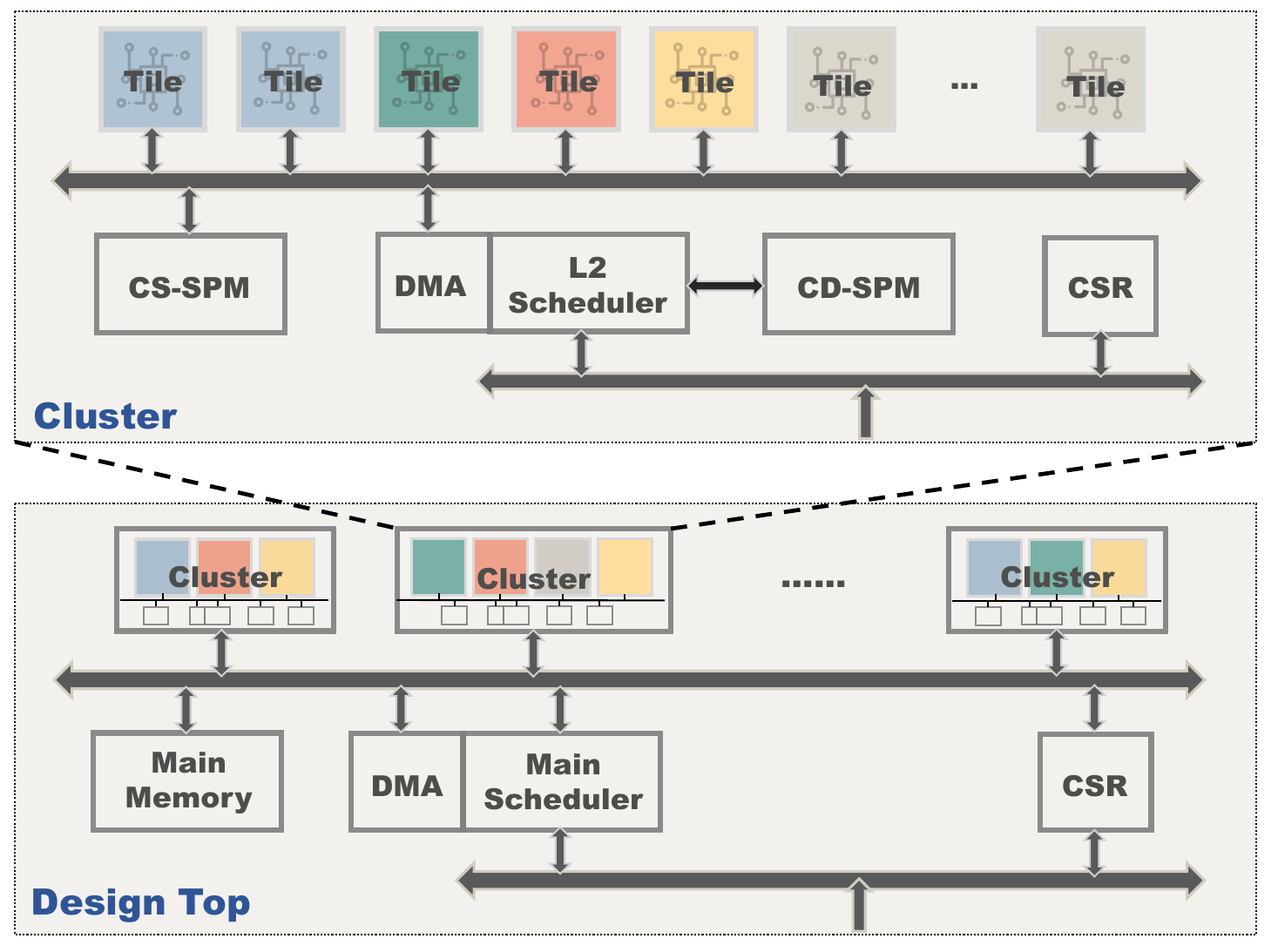}
  \caption{Overview of proposed design}
  \label{arch}
\end{figure}

\section{System Design}
\subsection{Architecture}
Figure \ref{arch} illustrates the proposed architecture for WBP. It is structured hierarchically, with a \textit{cluster} composed of several subordinate \textit{tiles}. Each tile is composed of a lightweight RISC-V scalar core, a customized vector extension unit (VXU), and a scratchpad memory (T-SPM). The scalar core is equipped with the RV32IM and custom vector instruction set architecture (ISA). The decode stage of the instruction pipeline identifies the instructions and forwards them to the tightly-coupled VXU through a queue when necessary. The VXU handles data in a SIMD fashion, wherein parallel lanes and element exchange engine (EXE) perform in-lane and inter-lane execution. The T-SPM stores the codes to be executed and the corresponding data, which will be further discussed in section \ref{Locality}. Additionally, control and status registers (CSRs) provide extra information about the tile.

At the next level of the hierarchy, i.e., the cluster, a level-two (L2) direct memory access (DMA) engine is used to orchestrate tiles. This is done by a scheduler, which is essentially a scalar processor with a scheduling scheme, that fetches instructions from a dedicated scratchpad memory (CD-SPM). Additionally, a shared memory (CS-SPM) is implemented in the cluster as a swap space for the tiles. The top level of the hierarchy is similar to the cluster level and consists of a main scheduler, a main DMA engine, and several clusters. The main scheduler is responsible for conducting high-level scheduling and instructing the main DMA engine to move codes and data from the main memory to the clusters. Further information about the scheduling procedure can be found in section \ref{multi-level}.

In addition to the classic manycore architecture, certain techniques are employed to guarantee the latency and effectiveness of the proposed design, as outlined below.

\begin{figure}[!t]
  \centering
  \subfloat[]{
    \includegraphics[width=0.95\linewidth]{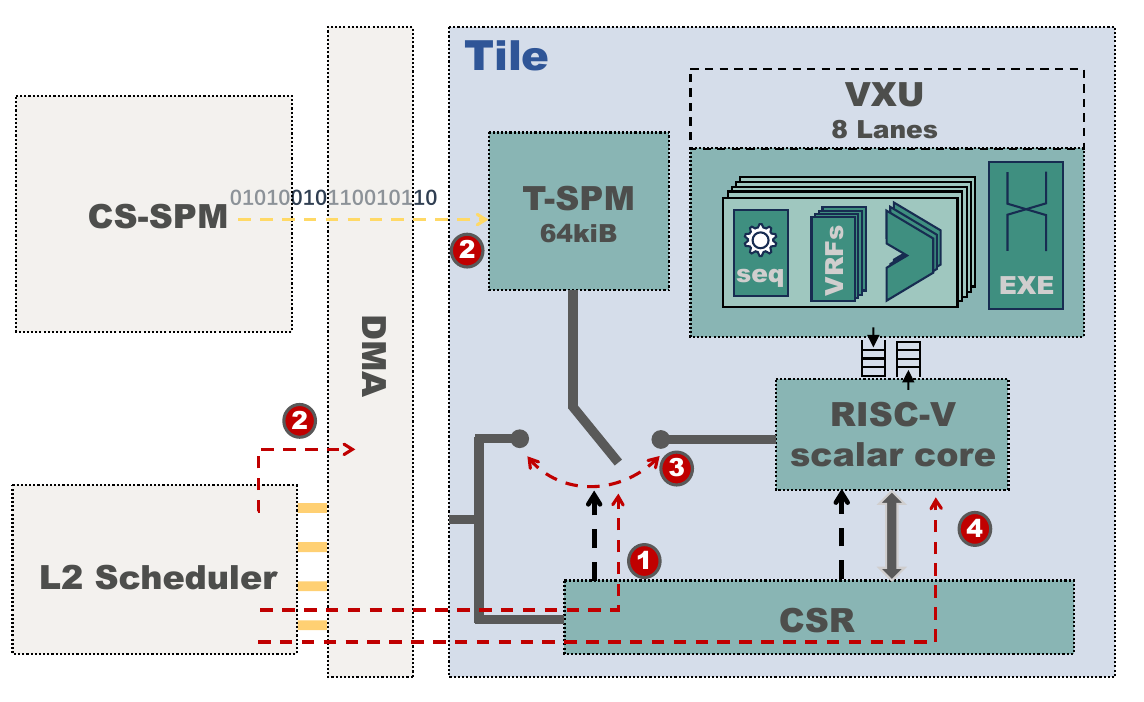}
    \label{pack-and-ship-before}
  }
  
  \subfloat[]{
    \includegraphics[width=0.95\linewidth]{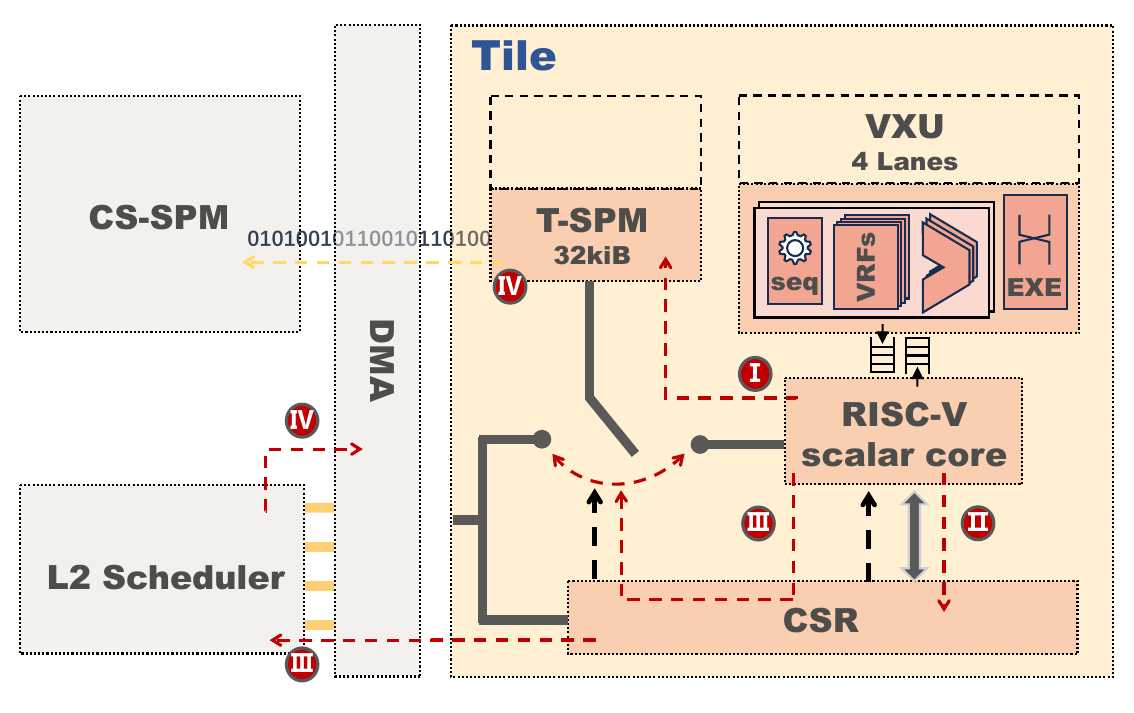}
    \label{pack-and-ship-after}
  }
  \caption{Detailed sequence of proposed \textit{pack-and-ship} scheme before and after the execution in a tile. Note that two tiles are heterogeneous.}
  \label{pack-and-ship}
  \vspace{-0.4cm}
\end{figure}

\subsubsection{Pack-and-ship NUMA architecture}
\label{Locality}
The system is designed with a non-uniform memory access (NUMA) approach to reduce the uncertainty and the large cache overhead in the memory hierarchy. Each tile and cluster has its own single-port private memory, i.e. T-SPM and CS-SPM, which can only be accessed by the \textit{outside} DMA and its own. In other words, the subordinate computation units view these private memories, rather than the main memory, as the entire memory space. Two factors make this possible. Firstly, the size of the instructions and the data of one module in WBP is relatively small, resulting in a moderate static random-access memory (SRAM) footprint. Secondly, the proposed NUMA architecture prevents all the computation units from randomly accessing the main memory. The DMA engine transfers data in the burst mode, which helps to hide the latency of the memory access.

Figures \ref{pack-and-ship-before} and \ref{pack-and-ship-after} illustrate the \textit{pack-and-ship} feature before and after its execution in two separate tiles. Before the process is carried out, the L2 scheduler and the DMA perform the following steps: (1) The scheduler first uses an atomic instruction to alter the port direction of the T-SPM to the bus matrix, which affects a CSR in the tile connected to multiplexers before the T-SPM; (2) The DMA is then programmed by the scheduler to move the codes and data from the CS-SPM to the predetermined location on the T-SPM; (3) The port direction of the T-SPM is then changed to the core inside the tile; (4) The scheduler executes another atomic instruction to deactivate the reset signal of the scalar core. At this point, the tile has been deployed and is ready to run.

The following steps occur after the execution: (I) The tile stores the register values in a predetermined spot on the T-SPM, which is specified in the tile and scheduler codes; (II) The tile notifies the number of return values by altering the CSRs in the tile; (III) The port direction is changed and an interrupt is sent to the scheduler to indicate the completion of the computation; (IV) After the notification, the scheduler orders the DMA to retrieve the data to the CS-SPM.

Similarly, the cluster level carries out data transfers between the main memory and CS-SPMs in a similar fashion, but with fewer CSR settings. The powerful locality of the proposed NUMA architecture reduced the data movement between tiles or clusters to reduce the memory wall and the activity of the interconnect.

\subsubsection{Heterogeneous configuration}
The architecture we have designed can be tailored to different wireless communication protocols by configuring tiles and clusters. This is done without the need for additional transistors. The following is a list of all the dimensions of configurations that our architecture supports.

\begin{itemize}
    \item \textit{Number of clusters and tiles:} These two dimensions increase the thread-level parallelism and task-level parallelism accordingly, so that transmit and receive chains in consecutive time slots as well as processing modules within the chains can be dealt with in parallel (Section \ref{sdf}). The throughput of protocols directly affects the scale of the architecture.
    \item \textit{SPM footprint:} The size of T-SPMs has an effect on different types of computations. For instance, the fast Fourier transform (FFT) requires a lot of multiply-accumulate operations. However, the intermediate variables (i.e. butterfly outputs) used in each butterfly phase have a short lifespan. This makes it a computationally intensive task. On the other hand, the belief propagation (BP) algorithm of polar codes is memory-intensive, since the intermediate log-likelihood ratios must be stored for the next iteration \cite{park20144}. The size of CS-SPMs is related to the multi-threading capability within one cluster (section \ref{multi-level}), which affects the tile utilization.
    \item \textit{Number of lanes and vector register files (VRFs) in VXU:} The DLP is increased by adding more lanes, and the maximum vector length capacity in runtime is increased by adding more VRFs. Further exploration of the VXU is to be done in the future.
\end{itemize}

\begin{algorithm}[!t]
\caption{The thread level scheduling scheme}\label{alg1}
\SetKwInput{KwInput}{Input}
\SetKwInput{KwOutput}{Output}
\SetKwProg{KwFunc}{Function}{}{}
\SetKw{KwIn}{in}
\SetKw{KwGoto}{goto}
\SetKwData{KwREADY}{READY}
\SetKwData{KwNULL}{NULL}
\SetKwData{KwTRUE}{TRUE}
\SetKwFunction{KwStatus}{status}
\SetKwFunction{KwDeployed}{codeDeployed}
\SetKwFunction{KwMalloc}{memalloc}
\SetKwFunction{KwMempack}{mempack}
\SetKwFunction{KwLRU}{getClusterLRU}
\SetKwFunction{KwThreadManager}{threadManager}
\SetKwFunction{KwThreadRegister}{threadRegistration}
\begin{small}
\KwInput{$tSet$, $cSet$ - Set of software threads and \emph{available} hardware clusters in the system, respectively}
\KwOutput{$aSet$ - Set of the beginning address of data and DAG ready to be transferred by DMA}
$aSet \leftarrow \varnothing$\;

\ForEach{$tID$ \KwIn $tSet$}
{
    $addr \leftarrow \varnothing$\;
    
    \If{$tID$.\KwStatus $=$ \KwREADY}
    {
        \tcc{Find a cluster that has already deployed the DAG before}
        $cID \leftarrow$ \KwDeployed{tID}\; 
        \If{$cID \neq $ \KwNULL}
        {
            \label{alg1:if_cid_not_null}
            \tcc{Memory request for the thread data}
            $addr \leftarrow$ \KwMalloc{$cID$, $tID$.data}\; \label{alg1:datamalloc_notnull}      
        }
        \Else
        {
            \tcc{Get a new physical cluster ID}
            \ForEach{$cID$ \KwIn $cSet$}
            {
                \tcc{Make an inquiry per cluster}
                $status \leftarrow$ \KwThreadManager{cID}\; \label{alg1:Threadmgr}
                \If{$status =$ \KwTRUE}
                {
                    \label{alg1:if_status_true}
                    \KwGoto line \textbf{\ref{alg1:mempack_null}}\; \label{alg1:foreach_break}
                }
            }
            \tcc{Drop the least recently used DAG}
            $cID \leftarrow$ \KwLRU{}\; \label{alg1:LRU}
            \tcc{Register SW thread to HW cluster}
            \KwThreadRegister{$cID$, $tID$}\; \label{alg1:thread_register}
            \tcc{Transfer DAG along with data}
            $payload \leftarrow$ \KwMempack{$tID$.data, $tID$.DAG}\; \label{alg1:mempack_null}
            $addr \leftarrow$ \KwMalloc{$cID$, payload}\; \label{alg1:datamalloc_null}
        }
        $aSet[tID] \leftarrow addr$
    }
}
\KwRet{$aSet$}
\end{small}
\end{algorithm}
\subsection{Execution Model}
\subsubsection{Dataflow model in WBP}
\label{sdf}
The dataflow model \cite{veen1986dataflow} is an ideal fit for WBP, as the modules in the flow are strongly linked in sequence, which can be represented as a dependency analysis graph (DAG). The subsequent module will only be activated when all the preceding modules have completed their tasks. Furthermore, the WBP follows a similar flow over time, which enhances data locality since the DAG information is unlikely to be reconfigured on the hardware.

We propose a multi-level dataflow model with \textit{task} and \textit{thread} granularity. A task typically represents a module, such as channel equalization, in the processing flow running on a tile. A thread is composed of several related tasks, which symbolizes a complete transmit or receive processing flow. The L2 and main schedulers are responsible for scheduling tasks and threads, respectively. Compared to traditional single-level scheduling, our proposed model is less likely to be \textit{scheduler-bounded}, meaning the scheduler is unable to allocate sufficient hardware resources due to various workloads. By deploying multiple schedulers at different levels, the overall scheduling capability is improved by reducing the pressure on the main scheduler.

\begin{figure}[!t]
  \centering
  \includegraphics[width=\linewidth]{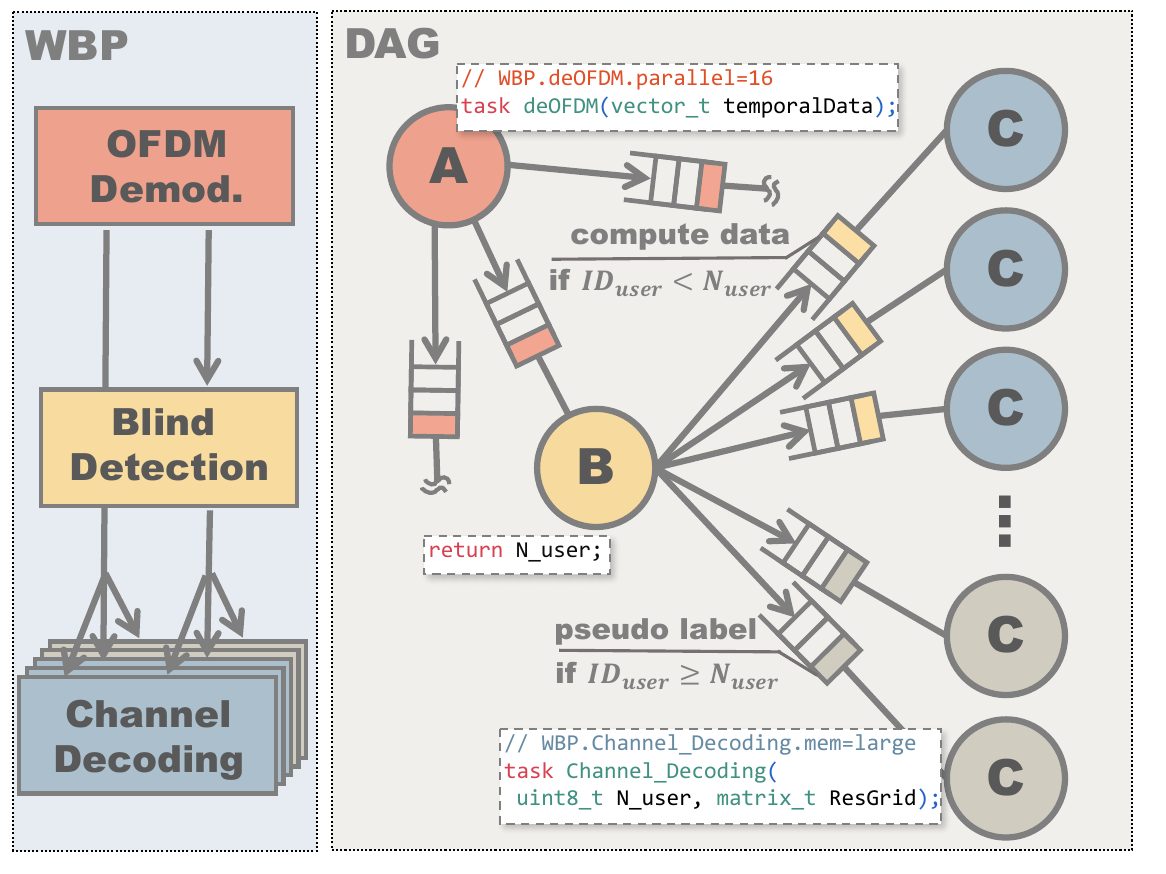}
  \caption{A partial example of WBP and corresponding DAG.}
  \label{dataflow}
\end{figure}

Our proposed dataflow model is tailored to the WBP by incorporating domain-specific features, as depicted in Figure \ref{dataflow}. Firstly, it allows for the identification of tasks with different workload characteristics. Attributes can be added to tasks to provide guidance to the schedulers about which type of tile is most suitable for deployment. This makes full use of the heterogeneous architecture by narrowing down the available tiles to those that meet the requirements. Secondly, our dataflow model enables the adjustment of DAGs at runtime. Unlike dynamic tasking in \cite{huang2021taskflow}, where a parent task is followed by a group of spawned tasks, tasks can be dismissed on-the-fly in our dataflow model after the worst-case DAG has been determined. For example, in the receive chain of the WBP, a blind detection task may be followed by a maximum of 20 channel decoding tasks, as the maximum possible number of users in a time slot of the resource grid is 20. However, if the return value of the blind detection task (i.e. the number of users detected) is much lower than the worst-case DAG, the computation burden can be significantly reduced due to the linear nature of the WBP, which avoids the exponential growth of tasks in the DAG.

\begin{figure}[!t]
  \centering
  \includegraphics[width=\linewidth]{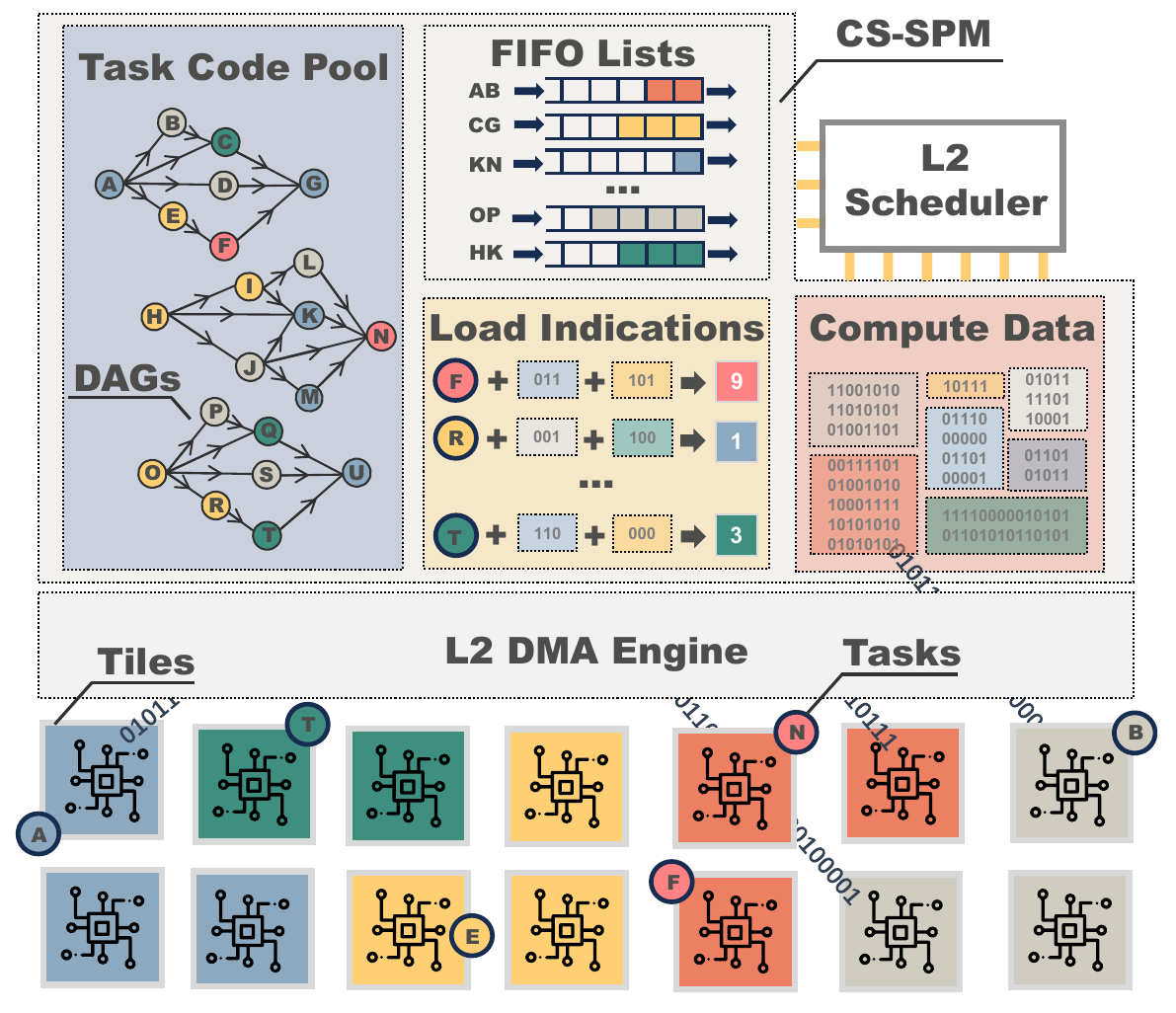}
  \caption{Tile-level scheduling scheme.}
  \label{L2_scheduler}
  \vspace{-0.6cm}
\end{figure}

\subsubsection{Multi-level scheduling scheme}
\label{multi-level}
The main scheduler and the L2 schedulers collaborate to carry out multi-level scheduling. The main scheduler is responsible for supervising clusters and delivering DAGs and data at the thread level. Algorithm \ref{alg1} outlines the framework of thread-level scheduling. We employ a \textit{lazy-deletion} mechanism to simplify thread management. When delivering a thread, the main scheduler records the thread ID and cluster ID in a table (line \ref{alg1:thread_register}). When the thread reaches the end of its life, the main scheduler does not free up the memory section of DAGs. If the main scheduler encounters data with the same DAG, it only transfers the data section by searching the table to check if the DAG has been deployed to a cluster before (line \ref{alg1:if_cid_not_null}-\ref{alg1:datamalloc_notnull}). Otherwise, the main scheduler orders the main DMA engine to gather both DAG and data to the CS-SPMs, replacing the original DAG with a least recently used (LRU) approach (line \ref{alg1:LRU}-\ref{alg1:datamalloc_null}). We observe that the low complexity of a DAG in a cluster results in the under-utilization of tiles, prompting us to support the multi-threading of a cluster. Before the thread registration, the scheduler inquires the thread manager of each cluster (line \ref{alg1:Threadmgr}). The thread manager is a low-cost hardware unit that monitors the status of each thread in the cluster. The system is more likely to be stable if there are sufficient tiles and clusters in the system.

After interacting with the main memory, the task-level scheduling carries out all the data movements within the chip. As shown in Figure \ref{L2_scheduler}, the CS-SPM is divided into four sections, which are accessed in sequence by the L2 scheduler and DMA engine. The scheduler first goes through the nodes (i.e. tasks) of a DAG in the \textit{task code pool}. During this process, the readiness of the node is checked by the status of the first-in-first-out (FIFO) queues in the \textit{FIFO lists}. These software FIFOs keep track of all the edges between the nodes of the DAGs. The letter pairs in Figure \ref{L2_scheduler} represent the source and sink nodes, respectively. The scheduling result, which is the pair of the task to be processed and the preferred tile determined by attributes, is stored in the \textit{load indication} section. Finally, the DMA engine, based on the load indication, transfers the data pointed by the FIFO lists from the \textit{compute data} section to the heterogeneous tiles, which are colored differently.

\begin{table}
  \caption{Details of the prototype system.}
  \label{linkconfig}
  \begin{threeparttable}[b]
  \begin{tabular}{rl}
    \toprule
    Module & Configuration \\
    \midrule
    Channel coding & Polar codes \\
    Rate-matching & Redundant version 0 \\
    Scrambling & Gold sequence \\
    Modulation & QPSK\tnote{$\ast$} \\
    OFDM modulation & 128 subcarriers \\
    Channel estimation & Least squares \\
    Channel equalization & Zero forcing \\
    Channel decoding & Min-sum BP \\
  \bottomrule
\end{tabular}
\begin{tablenotes}
    \footnotesize
    \item [$\ast$] Quadrature phase shift keying
\end{tablenotes}
\end{threeparttable}
\vspace{-0.4cm}
\end{table}

\section{Evaluation}
\subsection{Experimental Setup}
\begin{figure}[!t]
  \centering
  \includegraphics[width=\linewidth]{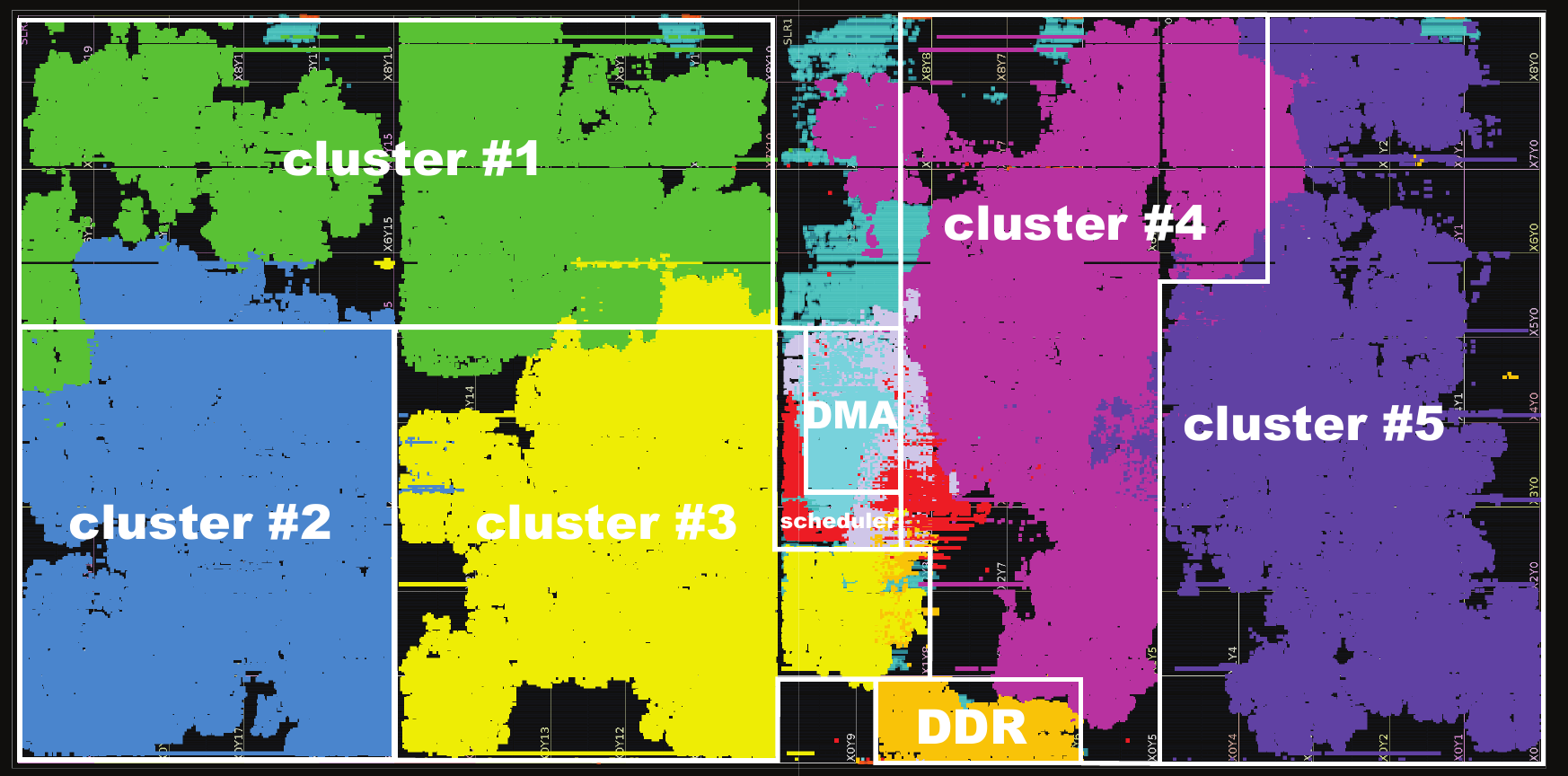}
  \caption{Layout of system with 5 clusters.}
  \label{floorplan}
\end{figure}

{\setlength{\parindent}{0pt}
\textbf{Methodology:} Our design is implemented in register transfer level (RTL) with System Verilog. Tiles and clusters can be configured easily by a high-level interactive script. We synthesize the design by Synopsys Design Compiler under SMIC 40 nm technology node. The target frequency is set to 500 MHz. 

\textbf{Task baselines:} We selected FFT and BP decoding of polar codes as examples of computationally and memory-intensive workloads. We compared the single-tile performance of our work with existing implementations on commercial platforms \cite{tidsp2008, chen20233} and hardware accelerators \cite{bertaccini2021buffer, park20144}. We used Synopsys VCS to measure the clock cycles of each tile. We normalized the throughput to per lane per GHz (64 lanes in our experiment) and the GPU to per CUDA core per GHz (128 CUDA cores per streamline multiprocessor in RTX 30 series) to make a fair comparison.

\textbf{Prototyping:} We present a simplified link layer as a proof of concept, which is summarized in Table \ref{linkconfig}. The number of clusters is set to 4 and 5, while the number of tiles ranges from 3 to 9. The tiles are divided into two types: the large (L) tile contains a 16-lane VXU with 32 VRFs, and the small (S) tile has an 8-lane VXU with 64 VRFs. The design is integrated with a 1600 MT/s double-data-rate (DDR) synchronous dynamic random access memory (SDRAM) and compiled by Synopsys ProtoCompiler. The system is brought up at a frequency of 30 MHz on a HAPS\textregistered-100 1F module with a Xilinx Ultrascale+ xcvu19p-fsva3824-1-e field programmable gate array (FPGA). An example layout is shown in Figure \ref{floorplan}.
}

\subsection{Experimental Results}
\subsubsection{Single-tile performance}

\begin{table}
  \caption{Single-tile performance compared with other works.}
  \label{single-tile}
  \begin{tabular}{c||ccc}
    \toprule
    Kernel                  & Platform                  & Config. & Performance\\
    \midrule
                            &                           & & Clock cycles\\
    \multirow{9}{*}{FFT}    & \multirow{3}{*}{DSP\cite{tidsp2008}} & $N=128$ & 588\\
                            &                                      & $N=512$ & 2559\\
                            &                                      & $N=2048$ & 11922\\
    \cline{2-4}
                            & \multirow{3}{*}{HW accel.\cite{bertaccini2021buffer}} & $N=128$ & 211\\
                            &                                                       & $N=512$ & 845\\
                            &                                                       & $N=2048$ & 3875\\
    \cline{2-4}
                            & \multirow{3}{*}{This work}& $N=128$ & 251\\
                            &                           & $N=512$ & 1122\\ 
                            &                           & $N=2048$ & 5073\\
    \hline
                                 &                      & &Norm. Thrpt.\\
    \multirow{5}{*}{BP decoding} & \multirow{2}{*}{GPU\cite{chen20233}}  & $N=512$ & 0.25\\
                                 &                                       & $N=1024$ & 0.21\\
    \cline{2-4}
                                 & ASIC\cite{park20144}                  & $N=1024$ & 15.23 \\
    \cline{2-4}
                                 & \multirow{2}{*}{This work}            & $N=512$ & 0.54\\
                                 &                                       & $N=1024$ & 0.53\\
  \bottomrule
\end{tabular}
\end{table}

As demonstrated in Table \ref{single-tile}, the performance of our tile lies between commercial hardware and application-specific integrated circuits (ASIC). It outperforms 2.3 times in FFT and 2 times in BP decoding compared to DSP and GPU. The primary motivation for our strategy of pack and ship is that it allows instructions and data to be preloaded into T-SPMs, thus decreasing the time taken for irregular memory requests. The fine-tuned DSP kernel fully utilizes eight function units, since load/store and arithmetic instructions are independent and can be packed into one VLIW. It is possible that our tile can keep up with a lower DLP by incorporating more domain-specific instructions. As for GPU, the performance of single-instruction multiple-threads (SIMT) is restricted by the slowest thread. Our tile design sticks to DLP at the tile level, while TLP is shifted to clusters. Nevertheless, limited by the common drawbacks of Von Neumann architecture, our design is still not competitive with customized hardware. The instruction-driven model requires a large amount of control to maintain its versatility, but no control is necessary for ASICs. Specifically, both \cite{bertaccini2021buffer} and \cite{park20144} have hardware designed for data reorganization.

\subsubsection{Ablation study}

\begin{table}
  \caption{Ablation study on thread-level improvements.}
  \label{ablation}
  \begin{tabular}{l|cc|cc}
    \toprule
    \multirow{2}{*}{\makecell[l]{Baseline\\ + extra features}} & \multicolumn{2}{c|}{Power (W)} & \multicolumn{2}{c}{Thrpt. (Mbps)}\\
    \cline{2-5}
    & 12T & 3C4T & Single-lvl. & Multi-lvl.\\
    \midrule
    12T / 3C4T arch.  &\multirow{3}{*}{3.24} &\multirow{3}{*}{3.45}    & 8.5 & 21.2\\
    + Multi-threading &                      &                         &64.1 & 84.7\\
    + Lazy-deletion   &                      &                         &68.5 & 93.6\\
  \bottomrule
\end{tabular}
\end{table}

We assess the power and throughput of a 3-cluster 2L2S-tile configuration (3C4T) based on Table \ref{linkconfig}. As a baseline, we also set up a 12-tile architecture (12T) with no thread-level scheduling. We acquire power data from Vivado reports after place and route, and only analyze dynamic power (without high-speed transceivers). Table \ref{ablation} reveals that, despite a 6.5\% power increase from extra scheduler cores and interconnect fabric, the throughput of our multi-level system is at least 1.3 times faster than the single-level system. It is noteworthy that the scheduler in 12T is not able to fully allocate tiles with multi-threading feature, which is approximately 9 tiles at maximum, while hierarchical schedulers in 3C4T eliminate this issue. Furthermore, the throughput additionally increases by 6.4\% and 9.5\% in 12T and 3C4T system under the lazy-deletion scheme.
\subsubsection{Link throughput experiment on prototype}

\begin{figure}[!t]
  \centering
  \includegraphics[width=\linewidth]{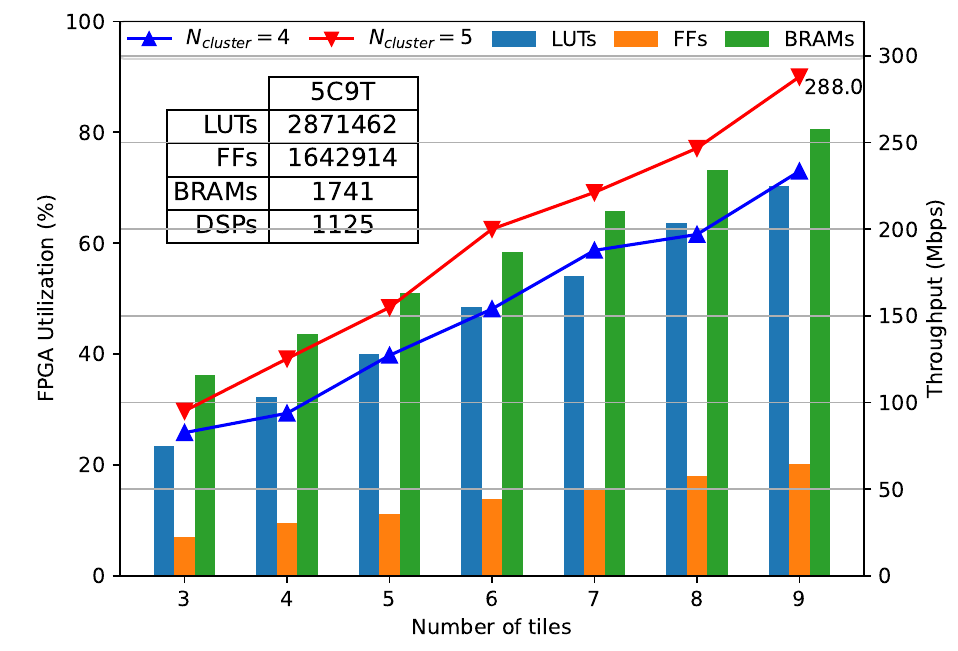}
  \caption{Throughput and hardware utilization under different configurations.}
  \label{performance}
\end{figure}

The graph in Figure \ref{performance} displays a linear increase in throughput and hardware resources. On average, throughputs with 5 clusters are 1.23 times higher than those with 4 clusters. The maximum throughput of a 5-cluster 9-tile system is 288 Mbps, as indicated in the upper right corner of the figure, which is sufficient to meet the throughput requirements of 5G under 16 carrier aggregation (maximum setting) and 5 downlink layers configuration in 5MHz bandwidth numerology. Although the greater throughput configuration cannot be immediately verified due to the limited hardware resources of the prototyping FPGA, the scalability of the proposed architecture is expected to be able to satisfy 5G or even more advanced wireless communication protocols. Additionally, the block RAMs utilization initially reaches the saturation threshold, which is the inevitable consequence of replacing frequent DDR access with extra distributed SRAM for "pre-fetching" (data-flow NUMA). To address the "SRAM starvation" issue present in the proposed architecture, more fine-grained heterogeneous strategies can be employed to customize the SPM footprint of each individual tile.

\section{Conclusion}
The cyclical and modular natures of WBP inspire us to develop a domain-specific architecture, as well as a novel execution model. In this paper, we propose a pack-and-ship approach within a cache-free NUMA system. Instructions and data are organized in bundles and delivered by schedulers to local scratchpad memory in order to reduce data movement costs. We also develop a hierarchical dataflow scheme along with two strategies, namely multi-threading and lazy-deletion, to exploit and allocate the hardware resources more efficiently. Based on extensive simulation and prototyping, our HW/SW co-design surpasses the existing architectures attributed to strong single-tile performance as well as flexible scalability and coarse-grained parallelism.

\begin{acks}
This work was supported by the National Natural Science Foundation of China (NSFC) under Grants 62271300, 12141107.
\end{acks}

\bibliographystyle{ACM-Reference-Format}
\bibliography{sample-base}


\begin{thebibliography}{21}


\ifx \showCODEN    \undefined \def \showCODEN     #1{\unskip}     \fi
\ifx \showDOI      \undefined \def \showDOI       #1{#1}\fi
\ifx \showISBNx    \undefined \def \showISBNx     #1{\unskip}     \fi
\ifx \showISBNxiii \undefined \def \showISBNxiii  #1{\unskip}     \fi
\ifx \showISSN     \undefined \def \showISSN      #1{\unskip}     \fi
\ifx \showLCCN     \undefined \def \showLCCN      #1{\unskip}     \fi
\ifx \shownote     \undefined \def \shownote      #1{#1}          \fi
\ifx \showarticletitle \undefined \def \showarticletitle #1{#1}   \fi
\ifx \showURL      \undefined \def \showURL       {\relax}        \fi
\providecommand\bibfield[2]{#2}
\providecommand\bibinfo[2]{#2}
\providecommand\natexlab[1]{#1}
\providecommand\showeprint[2][]{arXiv:#2}

\bibitem[Bertaccini et~al\mbox{.}(2021)]%
        {bertaccini2021buffer}
\bibfield{author}{\bibinfo{person}{Luca Bertaccini}, \bibinfo{person}{Luca Benini}, {and} \bibinfo{person}{Francesco Conti}.} \bibinfo{year}{2021}\natexlab{}.
\newblock \showarticletitle{To Buffer, or Not to Buffer? A Case Study on FFT Accelerators for Ultra-Low-Power Multicore Clusters}. In \bibinfo{booktitle}{\emph{2021 IEEE 32nd International Conference on Application-specific Systems, Architectures and Processors (ASAP)}}. IEEE, \bibinfo{pages}{1--8}.
\newblock


\bibitem[Bertuletti et~al\mbox{.}(2023)]%
        {bertuletti2023efficient}
\bibfield{author}{\bibinfo{person}{Marco Bertuletti}, \bibinfo{person}{Yichao Zhang}, \bibinfo{person}{Alessandro Vanelli-Coralli}, {and} \bibinfo{person}{Luca Benini}.} \bibinfo{year}{2023}\natexlab{}.
\newblock \showarticletitle{Efficient Parallelization of 5G-PUSCH on a Scalable RISC-V Many-Core Processor}. In \bibinfo{booktitle}{\emph{2023 Design, Automation \& Test in Europe Conference \& Exhibition (DATE)}}. IEEE, \bibinfo{pages}{1--6}.
\newblock


\bibitem[Chen et~al\mbox{.}(2022)]%
        {chen2022dxt501}
\bibfield{author}{\bibinfo{person}{Yang Chen}, \bibinfo{person}{Lin Liu}, \bibinfo{person}{Xuelin Feng}, {and} \bibinfo{person}{Jinglin Shi}.} \bibinfo{year}{2022}\natexlab{}.
\newblock \showarticletitle{DXT501: An SDR-Based Baseband MP-SoC for Multi-Protocol Industrial Wireless Communication}. In \bibinfo{booktitle}{\emph{2022 IEEE Symposium in Low-Power and High-Speed Chips (COOL CHIPS)}}. IEEE, \bibinfo{pages}{1--6}.
\newblock


\bibitem[Chen et~al\mbox{.}(2023)]%
        {chen20233}
\bibfield{author}{\bibinfo{person}{Yuxing Chen}, \bibinfo{person}{Xinyuan Qiao}, \bibinfo{person}{Keyue Deng}, \bibinfo{person}{Suwen Song}, {and} \bibinfo{person}{Zhongfeng Wang}.} \bibinfo{year}{2023}\natexlab{}.
\newblock \showarticletitle{3.8-Gbps Polar Belief Propagation Decoder on GPU}.
\newblock \bibinfo{journal}{\emph{IEEE Communications Letters}} (\bibinfo{year}{2023}).
\newblock


\bibitem[Codrescu et~al\mbox{.}(2014)]%
        {codrescu2014hexagon}
\bibfield{author}{\bibinfo{person}{Lucian Codrescu}, \bibinfo{person}{Willie Anderson}, \bibinfo{person}{Suresh Venkumanhanti}, \bibinfo{person}{Mao Zeng}, \bibinfo{person}{Erich Plondke}, \bibinfo{person}{Chris Koob}, \bibinfo{person}{Ajay Ingle}, \bibinfo{person}{Charles Tabony}, {and} \bibinfo{person}{Rick Maule}.} \bibinfo{year}{2014}\natexlab{}.
\newblock \showarticletitle{Hexagon DSP: An Architecture Optimized for Mobile Multimedia and Communications}.
\newblock \bibinfo{journal}{\emph{IEEE Micro}} \bibinfo{volume}{34}, \bibinfo{number}{2} (\bibinfo{year}{2014}), \bibinfo{pages}{34--43}.
\newblock


\bibitem[Ding et~al\mbox{.}(2020)]%
        {ding2020agora}
\bibfield{author}{\bibinfo{person}{Jian Ding}, \bibinfo{person}{Rahman Doost-Mohammady}, \bibinfo{person}{Anuj Kalia}, {and} \bibinfo{person}{Lin Zhong}.} \bibinfo{year}{2020}\natexlab{}.
\newblock \showarticletitle{Agora: Real-time Massive MIMO Baseband Processing in Software}. In \bibinfo{booktitle}{\emph{Proceedings of the 16th international conference on emerging networking experiments and technologies}}. \bibinfo{pages}{232--244}.
\newblock


\bibitem[Garcia-Saavedra and Costa-Perez(2021)]%
        {garcia2021ran}
\bibfield{author}{\bibinfo{person}{Andres Garcia-Saavedra} {and} \bibinfo{person}{Xavier Costa-Perez}.} \bibinfo{year}{2021}\natexlab{}.
\newblock \showarticletitle{O-RAN: Disrupting the Virtualized RAN Ecosystem}.
\newblock \bibinfo{journal}{\emph{IEEE Communications Standards Magazine}} \bibinfo{volume}{5}, \bibinfo{number}{4} (\bibinfo{year}{2021}), \bibinfo{pages}{96--103}.
\newblock


\bibitem[Huang et~al\mbox{.}(2021)]%
        {huang2021taskflow}
\bibfield{author}{\bibinfo{person}{Tsung-Wei Huang}, \bibinfo{person}{Dian-Lun Lin}, \bibinfo{person}{Chun-Xun Lin}, {and} \bibinfo{person}{Yibo Lin}.} \bibinfo{year}{2021}\natexlab{}.
\newblock \showarticletitle{Taskflow: A Lightweight Parallel and Heterogeneous Task Graph Computing System}.
\newblock \bibinfo{journal}{\emph{IEEE Transactions on Parallel and Distributed Systems}} \bibinfo{volume}{33}, \bibinfo{number}{6} (\bibinfo{year}{2021}), \bibinfo{pages}{1303--1320}.
\newblock


\bibitem[Inc.(2020)]%
        {ceva2020}
\bibfield{author}{\bibinfo{person}{CEVA{,} Inc.}} \bibinfo{year}{2020}\natexlab{}.
\newblock \bibinfo{booktitle}{\emph{Introducing CEVA-XC16}}.
\newblock
\urldef\tempurl%
\url{https://www.ceva-dsp.com/wp-content/uploads/2020/03/CEVA-XC16_introduction_non_nda_public-v2.pdf}
\showURL{%
\tempurl}


\bibitem[Instruments(2008)]%
        {tidsp2008}
\bibfield{author}{\bibinfo{person}{Texas Instruments}.} \bibinfo{year}{2008}\natexlab{}.
\newblock \bibinfo{booktitle}{\emph{TMS320C64x+ DSP Little-Endian DSP Library Programmer’s Reference}}.
\newblock
\urldef\tempurl%
\url{https://www.ti.com/lit/ug/sprueb8b/sprueb8b.pdf}
\showURL{%
\tempurl}


\bibitem[Jalier et~al\mbox{.}(2010)]%
        {jalier2010heterogeneous}
\bibfield{author}{\bibinfo{person}{Camille Jalier}, \bibinfo{person}{Didier Lattard}, \bibinfo{person}{Ahmed~Amine Jerraya}, \bibinfo{person}{Gilles Sassatelli}, \bibinfo{person}{Pascal Benoit}, {and} \bibinfo{person}{Lionel Torres}.} \bibinfo{year}{2010}\natexlab{}.
\newblock \showarticletitle{Heterogeneous vs Homogeneous MPSoC Approaches for a Mobile LTE Modem}. In \bibinfo{booktitle}{\emph{2010 Design, Automation \& Test in Europe Conference \& Exhibition (DATE 2010)}}. IEEE, \bibinfo{pages}{184--189}.
\newblock


\bibitem[Kelkar and Dick(2021)]%
        {kelkar2021nvidia}
\bibfield{author}{\bibinfo{person}{Anupa Kelkar} {and} \bibinfo{person}{Chris Dick}.} \bibinfo{year}{2021}\natexlab{}.
\newblock \showarticletitle{NVIDIA Aerial GPU Hosted AI-on-5G}. In \bibinfo{booktitle}{\emph{2021 IEEE 4th 5G World Forum (5GWF)}}. IEEE, \bibinfo{pages}{64--69}.
\newblock


\bibitem[Ling et~al\mbox{.}(2015)]%
        {ling2015macron}
\bibfield{author}{\bibinfo{person}{Xiang Ling}, \bibinfo{person}{Yiou Chen}, \bibinfo{person}{Zhiliang Yu}, \bibinfo{person}{Shihua Chen}, \bibinfo{person}{Xiaodong Wang}, {and} \bibinfo{person}{Gui Liang}.} \bibinfo{year}{2015}\natexlab{}.
\newblock \showarticletitle{MACRON: The NoC-based Many-Core Parallel Processing Platform and Its Applications in 4G Communication Systems}. In \bibinfo{booktitle}{\emph{2015 23rd Euromicro International Conference on Parallel, Distributed, and Network-Based Processing}}. IEEE, \bibinfo{pages}{396--403}.
\newblock


\bibitem[Mahurin(2023)]%
        {mahurin2023qualocmm}
\bibfield{author}{\bibinfo{person}{Eric Mahurin}.} \bibinfo{year}{2023}\natexlab{}.
\newblock \showarticletitle{Qualcomm{\textregistered} Hexagon™ NPU}. In \bibinfo{booktitle}{\emph{2023 IEEE Hot Chips 35 Symposium (HCS)}}. IEEE Computer Society, \bibinfo{pages}{1--19}.
\newblock


\bibitem[Park et~al\mbox{.}(2014)]%
        {park20144}
\bibfield{author}{\bibinfo{person}{Youn~Sung Park}, \bibinfo{person}{Yaoyu Tao}, \bibinfo{person}{Shuanghong Sun}, {and} \bibinfo{person}{Zhengya Zhang}.} \bibinfo{year}{2014}\natexlab{}.
\newblock \showarticletitle{A 4.68 Gb/s Belief Propagation Polar Decoder with Bit-Splitting Register File}. In \bibinfo{booktitle}{\emph{2014 Symposium on VLSI Circuits Digest of Technical Papers}}. IEEE, \bibinfo{pages}{1--2}.
\newblock


\bibitem[Schoeberl et~al\mbox{.}(2015)]%
        {schoeberl2015t}
\bibfield{author}{\bibinfo{person}{Martin Schoeberl}, \bibinfo{person}{Sahar Abbaspour}, \bibinfo{person}{Benny Akesson}, \bibinfo{person}{Neil Audsley}, \bibinfo{person}{Raffaele Capasso}, \bibinfo{person}{Jamie Garside}, \bibinfo{person}{Kees Goossens}, \bibinfo{person}{Sven Goossens}, \bibinfo{person}{Scott Hansen}, \bibinfo{person}{Reinhold Heckmann}, {et~al\mbox{.}}} \bibinfo{year}{2015}\natexlab{}.
\newblock \showarticletitle{T-CREST: Time-Predictable Multi-Core Architecture for Embedded Systems}.
\newblock \bibinfo{journal}{\emph{Journal of Systems Architecture}} \bibinfo{volume}{61}, \bibinfo{number}{9} (\bibinfo{year}{2015}), \bibinfo{pages}{449--471}.
\newblock


\bibitem[Stern(2013)]%
        {QorIQ2013}
\bibfield{author}{\bibinfo{person}{Barry Stern}.} \bibinfo{year}{2013}\natexlab{}.
\newblock \bibinfo{booktitle}{\emph{Next-Generation Wireless Network Bandwidth and Capacity Enabled by Heterogeneous and Distributed Networks White Paper}}.
\newblock
\urldef\tempurl%
\url{https://www.nxp.com/docs/en/white-paper/QORIQQONVERGEWP.pdf}
\showURL{%
\tempurl}


\bibitem[Tan et~al\mbox{.}(2011)]%
        {tan2011sora}
\bibfield{author}{\bibinfo{person}{Kun Tan}, \bibinfo{person}{He Liu}, \bibinfo{person}{Jiansong Zhang}, \bibinfo{person}{Yongguang Zhang}, \bibinfo{person}{Ji Fang}, {and} \bibinfo{person}{Geoffrey~M Voelker}.} \bibinfo{year}{2011}\natexlab{}.
\newblock \showarticletitle{Sora: High-Performance Software Radio Using General-Purpose Multi-Core Processors}.
\newblock \bibinfo{journal}{\emph{Commun. ACM}} \bibinfo{volume}{54}, \bibinfo{number}{1} (\bibinfo{year}{2011}), \bibinfo{pages}{99--107}.
\newblock


\bibitem[Veen(1986)]%
        {veen1986dataflow}
\bibfield{author}{\bibinfo{person}{Arthur~H Veen}.} \bibinfo{year}{1986}\natexlab{}.
\newblock \showarticletitle{Dataflow Machine Architecture}.
\newblock \bibinfo{journal}{\emph{ACM Computing Surveys (CSUR)}} \bibinfo{volume}{18}, \bibinfo{number}{4} (\bibinfo{year}{1986}), \bibinfo{pages}{365--396}.
\newblock


\bibitem[Venkataramani et~al\mbox{.}(2020)]%
        {venkataramani2020spectrum}
\bibfield{author}{\bibinfo{person}{Vanchinathan Venkataramani}, \bibinfo{person}{Aditi Kulkarni}, \bibinfo{person}{Tulika Mitra}, {and} \bibinfo{person}{Li-Shiuan Peh}.} \bibinfo{year}{2020}\natexlab{}.
\newblock \showarticletitle{SPECTRUM: A Software-Defined Predictable Many-Core Architecture for LTE/5G Baseband Processing}.
\newblock \bibinfo{journal}{\emph{ACM Transactions on Embedded Computing Systems (TECS)}} \bibinfo{volume}{19}, \bibinfo{number}{5} (\bibinfo{year}{2020}), \bibinfo{pages}{1--28}.
\newblock


\bibitem[Yang et~al\mbox{.}(2013)]%
        {yang2013bigstation}
\bibfield{author}{\bibinfo{person}{Qing Yang}, \bibinfo{person}{Xiaoxiao Li}, \bibinfo{person}{Hongyi Yao}, \bibinfo{person}{Ji Fang}, \bibinfo{person}{Kun Tan}, \bibinfo{person}{Wenjun Hu}, \bibinfo{person}{Jiansong Zhang}, {and} \bibinfo{person}{Yongguang Zhang}.} \bibinfo{year}{2013}\natexlab{}.
\newblock \showarticletitle{BigStation: Enabling Scalable Real-Time Signal Processingin Large MU-MIMO Systems}.
\newblock \bibinfo{journal}{\emph{ACM SIGCOMM Computer Communication Review}} \bibinfo{volume}{43}, \bibinfo{number}{4} (\bibinfo{year}{2013}), \bibinfo{pages}{399--410}.
\newblock


\end{thebibliography}










\end{document}